\documentclass[preprint]{elsarticle}

\usepackage[utf8]{inputenc}
\usepackage{hyperref}
\usepackage[english]{babel}

\usepackage{tabularx}
\usepackage{paralist}

\usepackage{amsfonts}
\usepackage{graphicx}
\usepackage{listings}
\lstdefinelanguage{Scala}{
  morekeywords={abstract,case,catch,class,def,%
    do,else,extends,false,final,finally,%
    for,if,implicit,import,match,mixin,%
    new,null,object,override,package,%
    private,protected,requires,return,sealed,%
    super,this,throw,trait,true,try,%
    type,val,var,while,with,yield},
  otherkeywords={=>,<-,<\%,<:,>:,\#,@},
  sensitive=true,
  morecomment=[l]{//},
  morecomment=[n]{/*}{*/},
  morestring=[b]",
  morestring=[b]',
  morestring=[b]"""
}

\usepackage[draft]{fixme}

\usepackage{listings}
\begin{document}

NOTICE: this is the author’s version of a work that was accepted for publication by Elsevier. Changes resulting from the publishing process, such as peer review, editing, corrections, structural formatting, and other quality control mechanisms may not be reflected in this document. Changes may have been made to this work since it was submitted for publication. A definitive version was subsequently published in Journal of Computational Science, Available online 29 July 2015, http://dx.doi.org/10.1016/j.jocs.2015.07.003 .

\title{Massively-concurrent Agent-based\\ Evolutionary Computing}

\author[agh]{D.~Krzywicki}
\ead{daniel.krzywicki@agh.edu.pl}

\author[agh]{W.~Turek}
\ead{wojciech.turek@agh.edu.pl}

\author[agh]{A.~Byrski\corref{cor1}}
\ead{olekb@agh.edu.pl}

\author[agh]{M.~Kisiel-Dorohinicki}
\ead{doroh@agh.edu.pl}

\cortext[cor1]{Corresponding author}

\address[agh]{AGH University of Science and Technology\\ Faculty of Computer Science, Electronics and Telecommunications}

\begin{abstract}
The fusion of the multi-agent paradigm with evolutionary computation
yielded promising results in many optimization problems.
Evolutionary multi-agent system (EMAS) are more similar to
biological evolution than classical evolutionary algorithms. However, technological limitations prevented the use of fully asynchronous agents in previous EMAS implementations. In this paper we present a new algorithm for agent-based evolutionary
computations. The individuals are represented as fully autonomous and asynchronous agents.
An efficient implementation of this algorithm was possible through the use of modern technologies
based on functional languages (namely Erlang and Scala), which natively support lightweight processes and asynchronous communication.
Our experiments show that such an asynchronous approach is both faster and more efficient in solving common optimization problems.
\end{abstract}


\maketitle

\section{Introduction}

Biological systems are asynchronous by nature. This fact is not always considered in biologically-inspired computing methods (e.g. metaheuristics, such as evolutionary algorithms \cite{michalewicz}). These systems usually use notions such as ``discrete'' generations, loosing such concepts as parallel ontogenesis,
or lack of global control. Nevertheless, such computing systems have proven to be effective in different optimization
problems. Moreover, some of them can be mathematically proven to work in terms of asymptotic stochastic guarantee of success (cf.\ works of Vose on simple genetic algorithm \cite{vose}). 

Agent-oriented systems should also be asynchronous by nature, as they are inspired by social or biological systems. 
Over the last decade, our group has worked on the design and development of decentralized evolutionary computations~\cite{ker} in the form of Evolutionary Multi-Agent Systems \cite{kc}. 
EMAS is a hybrid meta-heuristic which combines multi-agent systems with evolutionary algorithms. 
A dedicated mathematical formalism, based on Markov chains (similar to Vose's approach) was constructed and analysed \cite{biul1}, showing that EMAS may be also treated as a general-purpose optimization system. Besides that, a number of other formalisms along with dedicated frameworks 
implemented in different programming languages (like Java, Scala or Python) were developed
(see, e.g. \cite{facilitating,hgscamwa,cecmoj}).

The concept of hybridization of agent-based systems with evolutionary techniques can be  implemented in different ways, especially with regard to asynchronicity and concurrency, as well as distribution and parallelism.
There is a number of  popular agent-oriented frameworks which offer asynchronously communicating agents (such as Jadex \cite{Pokahr2013}, Jade \cite{Bellifemine2001} or MadKit\cite{Gutknecht2001}). However, they all share similar properties, such as heavyweight agents, at least partial FIPA-compliancy (JADE) or a BDI model (Jadex). It is also common for each agent to be executed as a
separate thread (e.g.\ in JADE). These traits are indeed appropriate to model flexible, coarse-grained, open systems. However, evidence suggest they are not best suited for closed systems with homogeneous agents
nor for fine-grained concurrency with large numbers of lightweight agents, which are both common in biologically-inspired population-based computing systems \cite{turek2013erlang}.

Therefore, dedicated tools for the above-mentioned class of agent-based systems have been constructed over the last 15 years. 
One of the successful implementations is the AgE platform\footnote{http://age.agh.edu.pl}, which supports phase-model and hybrid concurrency features (parts of the system are concurrent and parts are implemented as sequential processes). 
The AgE platform also has other advantages, such as significant support for reuse and flexible configuration management. 
Dedicated AgE implementations were constructed using Java, .NET and Python technologies.

However, the renewed interest in functional programming and languages such as Erlang and Scala brought new possibilities in terms of concurrent programming support. In a recent paper \cite{Krzywicki20141068}, we proposed a promising new approach to these kinds of agent-based algorithms. We used Erlang lightweight processes to implement a fined grained multi-agent system and bring more asynchronicity into usually synchronously implemented agent actions and interactions.  

In this paper, we present a significant progress over the research presented in \cite{Krzywicki20141068}, by extending our experiments to a Scala-based
implementation (in addition to the Erlang-based one) and comparing these two approaches.
The previous Erlang asynchronous implementation showed a small but statistically significant improvement in terms of efficiency, compared to a synchronous version. The results for the new Scala implementation provide much stronger evidence for the superiority of a massively-concurrent EMAS implementation over a traditional, synchronous one.

In the next sections we present the current state-of-the-art and introduce concepts of evolutionary multi-agent systems. Then, we describe the implementation of the synchronous and asynchronous versions of our algorithm, followed by our experimental settings and results. We end with a discussion of our results and conclude the paper with possible opportunities for future work.
                                    
\section{Large-scale Agent-based Systems}

The development of the software agent concepts and the theory of multi-agent systems took place in the last decades of the 20th century. At this point in time, the software engineering domain was strongly focused on the popularization of the object-oriented paradigm. As a result, the majority of agent systems and platforms for agent systems development was based on imperative languages with shared memory. This approach is in opposition to the assumptions of agent systems, which are based on the concept of message passing, communication between autonomous execution threads and do not allow any explicit shared state. 

Implementations of message passing concurrency in object oriented technologies resulted in significant limitations in both scale and performance of the developed solutions. The evaluation presented in \cite{turek2013erlang} shows that the most popular agent development platforms (Jade and Magentix) are limited to several thousands of simultaneous agents on a single computer. The limit was caused by the method used for implementing concurrently executing code of agents -- each agent required a separate operating system thread. This situation inhibited the development of large scale multi-agent systems for long time. 

Current trends in the domain of programming languages development focus on the integration of concepts from different paradigms and on the development of new languages dedicated for particular purposes. The renewed interest in the functional paradigm seems very significant in the context of agent systems development. The agent system for human population simulation, presented in \cite{FrankSteffenRaubal2001}, was implemented in Haskell. The authors emphasize that the source code was very short in respect to the complex functionality of the system.
The discussion presented in \cite{GrigoreCollier2011} focuses on the language features of Haskell in the context of agent systems. The authors show the usefulness of algebraic data types, roles and sessions in implementation of multi-agent algorithms. 

The popularization of the functional paradigm concepts is mostly caused by difficulties in efficient usage of multi-core CPUs in languages implementing a shared-memory concurrency model. The need of synchronization of all the operations on shared memory effectively prevents the applications from scaling on higher numbers of cores. The issue does not exist in the message passing concurrency model, which allows massively concurrent applications to run effectively on parallel hardware architectures. This fact triggered the development of languages and runtime environments which efficiently implement the message passing concurrency model. There are currently two major technologies of this kind being successfully used in industrial applications: Erlang and Scala with the Akka library. 

The concurrency model implemented by these technologies is based on the same assumptions as in the case of software agents. Therefore, these technologies are a very good basis for large scale and high performance multi-agent systems. Recent example of a Scala-based implementation of a custom multi-agent architecture can be found in \cite{ManateMunteanu2013}, where the authors present a system capable of processing and storing a large amount of messages gathered from sensing different devices.

Erlang technology has been found useful in this kind of applications much earlier. In 2003 the first agent development platform, called eXAT (\emph{erlang eXperimental Agent Tool}), has been presented in \cite{di2003exat}. The goal of this platform was to test the feasibility of using functional programming languages as a development tool for FIPA-compliant agents. An agent in eXAT is an actor-based, independent entity composed of \emph{behaviours}, which represent the functionality of an agent as a finite-state machine. Transitions between states are triggered by changes in the knowledge-base facts or by external messages. The original version of eXAT does not support agent migrations, however there is a version supporting this functionality \cite{Piotrowski2013}.  

eXAT platform overcomes the basic limitations of Java-based solutions. eXAT agents are based on Erlang lightweight processes, which can be created in millions on a single computer. Although the platform never became a mainstream tool, it should be noticed that it was the first environment which allowed to test the behaviour of large scale systems with truly parallel agents. 

Recent years brought different solutions based on Erlang technology, like the eJason system \cite{DiazEarle2013}. eJason is an Erlang implementation of Jason, which is a platform for the development of multi-agent systems developed in Java. The reason for rewriting the Jason platform in Erlang, pointed by the authors, are significant similarities between Jason agents and Erlang processes and the high performance and scalability of Erlang processes implementation. 

The ability to build and test massively-concurrent agent-based systems opens new possibilities of research in this domain. The algorithm presented in this paper is made possible by the high-performance implementations of a message-passing concurrency model offered by Erlang and Scala. 

\section{Evolutionary Multi-Agent Systems (EMAS)}

Generally speaking, evolutionary algorithms are usually perceived as universal optimization-capable metaheuristics (cf. theory of Vose \cite{vose}). 
However, the classical designs of evolutionary algorithms
(such as simple genetic algorithm \cite{goldberg}, evolution strategies etc. \cite{schwefel}) assume important
simplifications of the underlying biological phenomena. 
Such simplification mainly consists in avoiding direct implementation of such phenomena
observed in real-world biological systems, as 
dynamically changing environmental conditions, a dependency on multiple criteria,
the co-evolution of species, the evolution of the genotype-fenotype mapping, the assumption of neither global knowledge nor generational synchronization.

Of course, it does not mean that they are wrong per-se, as they have clearly proven themselves in solving difficult problems. However, there is still room for improvement, and the No Free Lunch Theorem \cite{wolpert} reminds us that the search for new optimization techniques will always be necessary.

One of the important drawback of classical evolutionary techniques is that they work on a number of data structures (\emph{populations}) and repeat in cycles (\emph{generations}) the same process of selecting parents and producing offspring using variation operators. 
Such an approach makes it difficult to implement large-scale, parallel implementations of evolutionary algorithms. Only trivial approaches to the parallelization of such algorithms were proposed (e.g. the master-slave model or parallel evolutionary algorithm \cite{cantu-paz1998survey}).

For the over 10 years, our group tried to overcome some of these limitations
by working on the idea of decentralised evolutionary computations~\cite{ker}, namely  Evolutionary Multi-agent Systems (EMAS) \cite{kc}. 
EMAS is a hybrid meta-heuristics which combines multi-agent systems with evolutionary algorithms. The basic idea of EMAS consists in evolving a population of agents (containing  the potential solutions to the problem in the form of genotypes). The agents are capable of doing different actions, communicating among themselves and with the environment, in order to find the optimal solution of the optimization problem.

According to classic definitions (cf. e.g., \cite{JenSycWoo:AAMAS98}) of a multi-agent system, there should not be global knowledge shared by all of agents. 
They should remain autonomous without the need to create any central authorities.
Therefore, evolutionary mechanisms such as selection needs to be decentralized, in contrast with traditional evolutionary algorithms. 
Using agent terminology, one can say that selective pressure is required to \emph{emerge} from peer to peer interactions between agents instead of being globally-driven.

\begin{figure}[h!] \centering
	\includegraphics[width=.7\columnwidth]{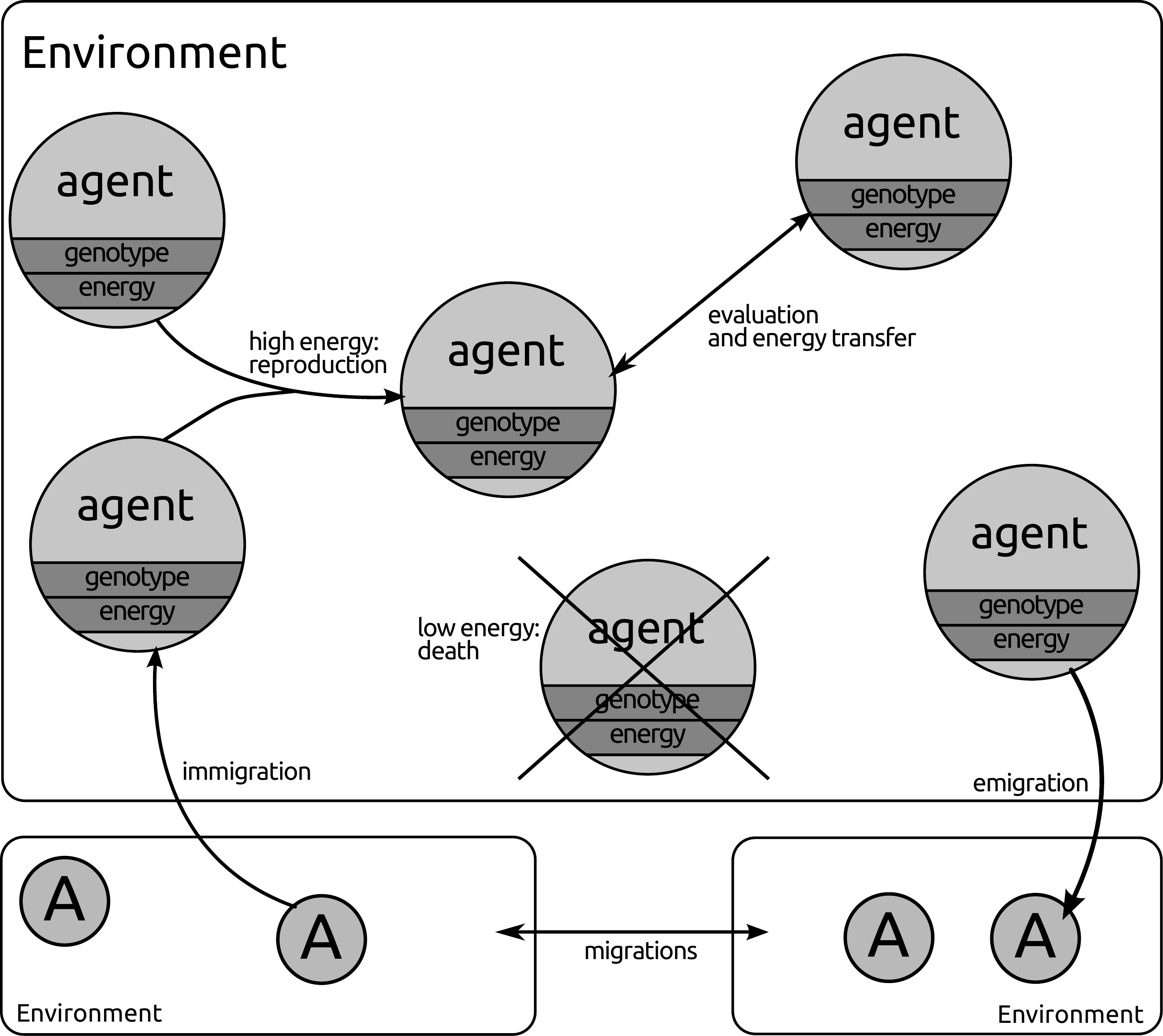}
	\caption{Structure and behaviour of EMAS operation.} \label{fig:emas}
\end{figure}

Thus, selection in EMAS is achieved by introducing a non-renewable resource, called life-energy. 
Agents receive part of the energy when they are introduced in the system. They exchange  energy based on the quality of their solution to the problem: worse agents move part of their energy to the better ones. The agents reaching certain energy threshold may reproduce, while the ones with low amount of energy die and are removed from the system. We show the principle of these operation in Fig. \ref{fig:emas}. For more details on EMAS design refer to \cite{ker}. 

Up till now, different EMAS implementations were applied to different problems (global, multi-criteria and multi-modal optimization in continuous and discrete spaces), and the results clearly showed superior performance in comparison to classical approaches (see \cite{ker}). 
It is to note, that besides acclaimed benchmark problems, as multi-modal functions
\cite{pisarskirugala,csci2013,korczynski}, and discrete benchmarks with 
clear practical application, like Low Autocorrelation Binary Sequence or Golomb Ruler problem \cite{DBLP:conf/iccS/KowolBK14,DBLP:conf/ecms/KolybaczKLBK13}
EMAS was also successfully applied to solving selected inverse problems
\cite{DBLP:journals/aghcs/WrobelTPB13,polnik}
leveraging its capability of quite low computational cost evaluated as number
of fitness function calls, compared to other classic approaches.

During the last years, we made also several approaches to construct efficient software targeted at variants of EMAS and at agent-based computing in general. 
We implemented dedicated tools in order to prepare fully-fledged frameworks convenient for EMAS computing (and several other purposes, as agent-based simulation). 
First, we focused on implementing decentralized agent behaviour, and the outcome were several fully synchronous versions, resulting in the implementation of the AgE platform\footnote{http://age.agh.edu.pl}. 
In this implementation we applied a phase-model of simulation, efficiently implementing such EMAS aspects as decentralized selection. We also supported the user with different component-oriented utilities, increasing reuse possibilities and allowing flexible configuration of the computing system. 

In a recent paper \cite{Krzywicki20141068}, we have presented a promising new approach to these kinds of algorithms.
We used Erlang lightweight processes to implement a fine-grained multi-agent system. 
Agents are fully asynchronous and autonomous in fulfilling their goals, such as exchanging resources with others, reproducing or being removed from the system. Agents are able to coordinate their behaviour with the use of mediating entities called meeting arenas.

\begin{figure}[h!] \centering
	\includegraphics[width=\columnwidth]{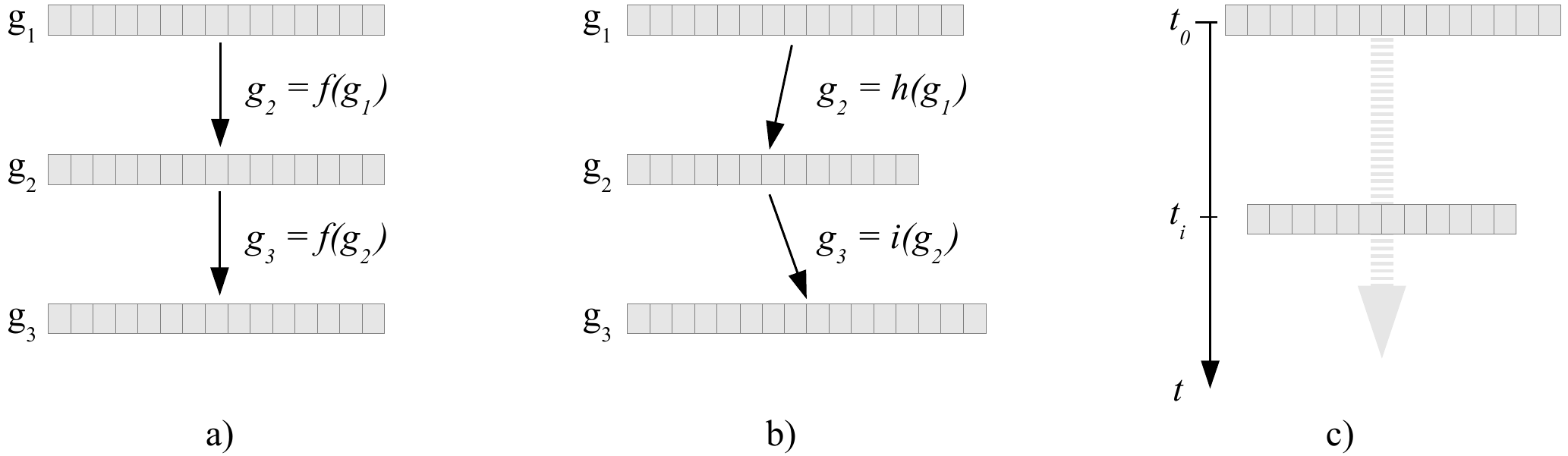}
	\caption{Concept of generation-oriented and generation-free evolutionary
		computing.} \label{fig:bestest}
\end{figure}

This approach brings us closer to the biological origins of evolutionary algorithms by removing artificial generations imposed by step-based implementations.  We show the concept of generation-oriented and generation-free computing in Fig.~\ref{fig:bestest}.

The first case (\emph{a}) shows the classical approach, consisting in
transforming a population of individuals by applying  a stochastic state transition function. This function is usually composed of some predefined operators, such as selection, crossover and mutation. 
The second case (\emph{b}) shows the EMAS approach (as it was realized in AgE platform), where different transformation functions are applied as the results of agent actions. This model still assumes the existence of generations, but on a technical level, as a result of a step-based simulation. 
In fact, both above cases use discrete-time based simulation. 

In contrast, the third case (\emph{c}) is a nearly continuous-time simulation (if we disregard the discrete nature of the machine itself). In this model, all agents may initiate actions at any possible time. The process scheduler makes sure they are given computing resources when they need them.

\section{Massively-concurrent EMAS implementation}
\label{sec:design}

The system presented in this work has been implemented in Erlang and Scala, as the lightweight concurrency model provided by these technologies is well suited for creating large-scale multi-agent systems. The implementation focuses on comparing different computational models in terms of their features and efficiency. The rest of this section describes the algorithms used in our evolutionary multi-agent system, the model of agents interactions and different implementations.

\subsection{Principle of system operation}

Every agent in the system is characterized by a vector of real values representing potential solution to the optimization problem. The vector is used for calculating the corresponding fitness value. The process of calculating the fitness value for a given solution is the most expensive operation. It is executed each time a new solution is generated in the system. 

Emergent selective pressure is achieved by giving agents a piece of non-renewable resource, called energy \cite{ker}. An initial amount of energy is given to a newly created agent by its parents. If the energy of two agents is below a required threshold, they fight by comparing their fitness value -- the better agent takes energy from the worse one. If an agent looses all its energy, it is removed from the system. Agents with enough energy reproduce and yield new agents. The genotype of the children is derived from their parents using variation operators. The number of agents may vary over time, however the system remains stable as the total energy remains constant.

As in the case of other evolutionary algorithms, the population of agents can be split into separated sub-populations. This approach helps preserving population diversity by introducing allopatric speciation and can also simplify parallel execution of the algorithm. In our case the sub-populations are called \emph{islands}. Information can be exchanged between the islands through agent migrations.

\subsection{Arenas}

An efficient implementation of meetings between agents is crucial for the overall performance of the algorithm. The meetings model has a significant impact on the properties of the algorithm, on its computational efficiency and on its potential for parallel execution.

A general and simple way to perform meetings is to shuffle the list of agents and then process pairs of agents sequentially or in parallel. However, this approach has several limitations:
\begin{compactitem}
	\item The whole population must be collected in order to shuffle agents and form the pairs. This approach is inappropriate in an algorithm which should be decentralized by nature.
	\item Agents willing to perform different actions can be grouped together -- all combinations of possible behaviours must be handled.
\end{compactitem}

\begin{figure}[h!]
\centering
\includegraphics[width=.8\columnwidth]{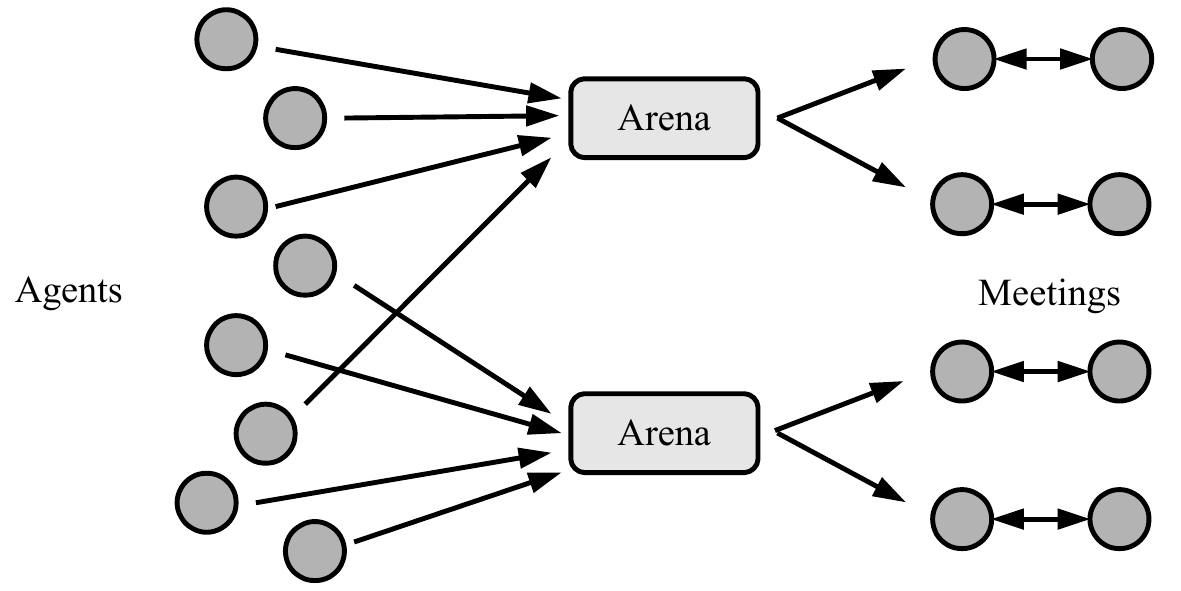}
\caption{Meeting arenas group similar agents and coordinate meetings between them.}
\label{fig:meetingArena}
\end{figure}

In our previous work \cite{Krzywicki2014}, a different approach was proposed. We proposed to group agents willing to perform the same action in dedicated \emph{meeting arenas}, following the Mediator design pattern. Every agent enters a selected arena depending on its amount of energy. Arenas split incoming agents into groups and trigger the actual meetings (see Fig. \ref{fig:meetingArena}). Each kind of agent behaviour is represented by a separate arena.

Therefore, the dynamics of the multi-agent system are fully defined by functions. The first function represents agent behaviour, which chooses an arena for each agent. The second function represents the meeting operation which is applied in every arena. 

This approach is similar to the MapReduce model, where arena selection corresponds to mapping and meeting logic to reduce operation. The pattern is very flexible, as it can be implemented in both a centralized and synchronous way or a decentralized and asynchronous one.

\subsection{Sequential implementation} 

The sequential version of the presented multi-agent system is implemented as a discrete event simulation. In each step the \emph{behaviour} function (see Listing~\ref{ls:agentBehaviour}) divides the population of agents into groups corresponding to available arenas.

\begin{figure}[h!]
\begin{lstlisting}[language=scala, label=ls:agentBehaviour, 
caption={In every step, agents choose an arena based on their current state}]
def behaviour(a: Agent) = a.energy match {
   0 => death
   x if x > 10 => reproduction
   x => fight
}
\end{lstlisting}
\end{figure}

Agents are first grouped according to their chosen arena and then a meeting function is applied on each such partition of the population. The partitions can be further subdivided into pairs of meeting agents and processed by applying a different meeting function which depends on the type of the arena (see Listing \ref{ls:arenas}). Every meeting results in a group of agents. The group can contain some new agents created as a result of reproduction and some with their state changed (e.g. by transferring energy). Some agents may be removed from the group if their energy equals $0$. Resulting groups are merged in order to form the new population, which is randomly shuffled before the next step.

\begin{figure}[h!]
\begin{lstlisting}[language=scala, label=ls:arenas, 
caption={Depending on the type of the arena, a different meeting happens, which transforms the incoming subpopulation of agents. The death arena simply return an empty sequence. Other arenas shuffle incoming agents, group them into pairs and apply a binary operator on every pair, concatenating results.}]
def meeting(arena: Arena, agents: Seq[Agent]) = 
   arena match {
      death => 
         Seq.empty[Agent]
      reproduction =>
         agents
            .shuffle
            .grouped(2)
            .flatMap(doReproduce)
      fight =>
         agents
            .shuffle
            .grouped(2)
            .flatMap(doFight)
   }
\end{lstlisting}
\end{figure}

If several islands are considered, each is represented as separate lists of agents. Migration between islands is performed at the end of each step by moving some agents between lists.

\subsection{Hybrid implementation} 

The introduction of coarse-grained concurrency in such a multi-agent system is rather straightforward. In our second implementation every island is assigned to a separate Erlang process/Scala actor responsible for executing the loop of the sequential algorithm described above. Islands communicate through message-passing, no other synchronization is needed.

The most significant difference regards agent migration. The behaviour function from Listing~\ref{ls:agentBehaviour} is modified by adding a migration action which is chosen with some fixed, low probability. The migration process is performed by a dedicated migration arena present on every island.

The migration arena removes agent from the local population and forwards it agent to a selected island chosen according to some topology and migration strategy. In every step the processes responsible for executing islands loop incorporates the incoming agents into their population.

\subsection{Concurrent implementation}

The first two implementation presented so far do not require massively concurrent execution, as the agents are represented as data structures processed sequentially by islands and arenas. Such approach does not reflect the autonomy of entities in the population and the true dynamics of relation between the entities, as each agent is forced to perform exactly one operation in each step of the algorithm. 

In order to achieve asynchronous behaviours of agents in the population, every agent and every arena has been implemented as a separate process/actor which communicates with the outside world only through message passing. The algorithm becomes fully asynchronous, as every agent acts at its own pace and there is no population-wide step. Meeting arenas are especially useful in this implementation, as they greatly simplify communication protocols. 

The algorithm of each agent is relatively simple. Depending on its current energy, every agent selects the action to perform and sends a message to the appropriate arena. Afterwards it waits for a message with the results of the meeting. 

As soon as enough agents gather in an arena, a meeting is triggered. As a result, new agents may be created and existing agents may be killed or replied with a message containing their new state.

Islands are logically defined as distinct sets of arenas. Each agent knows the addresses of all the arenas defining a single island, therefore it can only meet with other agents sharing the same set of arenas. Fights and reproductions arenas behave just as described in the previous version of the algorithm.

This migration process is greatly simplified in this version. Migrating an agent simply means changing the arenas it meets on. Migration arenas choose an island according to specified topology and send the addresses of the corresponding arenas back to the agent. The agent updates the set of arenas available to itself and resumes its behaviour. As it will now be able to meet with a different set of agents, it has indeed migrated.

\bigskip

The implementations in Erlang and Scala are relatively similar, as both realize exactly the same algorithms. In case of first two approaches (sequential and hybrid) no differences in behaviour of the implemented system was expected. On the other hand the execution of massively concurrent version can be significantly dependent on scheduling mechanisms implemented by the underlying process management system. The time of activities performed by each agent depends on the provided CPU access. 

The details of process scheduling mechanisms implemented by Erlang and Scala (Akka) are slightly different. Erlang was designed as a soft real-time platform, which means that each a process can be preempted in every moment in time, and none can claim more computational time than the others. The model implemented by Scala is more suited for reactive, message-driven programming, where a process can use CPU until it finished processing current messages. These tiny differences can influence the overall performance of the implemented systems but also can result in different behaviour of the populations.

Another difference is memory management. In Erlang, every process owns a separate stack and the content of messages needs to be copied between process memories. Scala uses the Java Memory Model and adds a message-passing layer on top of shared memory. The Akka actor library follows the principle "going from remote to local by way of optimization", therefore communication within a single Java Virtual Machine is very memory efficient, as immutable messages are safe to be shared between the sender and receiver.

\section{Methodology and Results}
\label{sec:results}
We used our multi-agent system to minimize the Rastrigin function, a common benchmarking function used to compare evolutionary algorithms. This function is highly multimodal with many local minima and one global minimum equal 0 at $\bar{x} = 0$. We used a problem size (the dimension of the function) equal to 100, in a domain equal to the hypercube $[-50, 50]^{100}$. 

The simulations were run on Intel Xeon X5650 nodes provided by the Pl-Grid\footnote{http://www.plgrid.pl/en} infrastructure at the ACC Cyfronet AGH\footnote{http://www.cyfronet.krakow.pl/en/}. We used up to 12 cores and 1 GB of memory. 

We tested the alternative approaches described in the previous section, implemented in both Erlang ans Scala. The experiments for hybrid and concurrent models were run on 1,2,4,8 and 12 cores. A sequential version in each language was also run on 2 cores (the second core being used for logging an management) - more cores did not improve the results.

Every experiment lasted 15 minutes and was repeated 30 times in order to obtain statistically significant results. The results further below are averaged over these 30 runs.

The hybrid model does not benefit from a number of cores higher than the number of islands. As we had 12 cores at our disposal, we used 12 islands in every experiment. Migration destinations were chosen at random in a fully connected topology.

\begin{figure}[!h]
    \centering 
    \includegraphics[width=\textwidth]{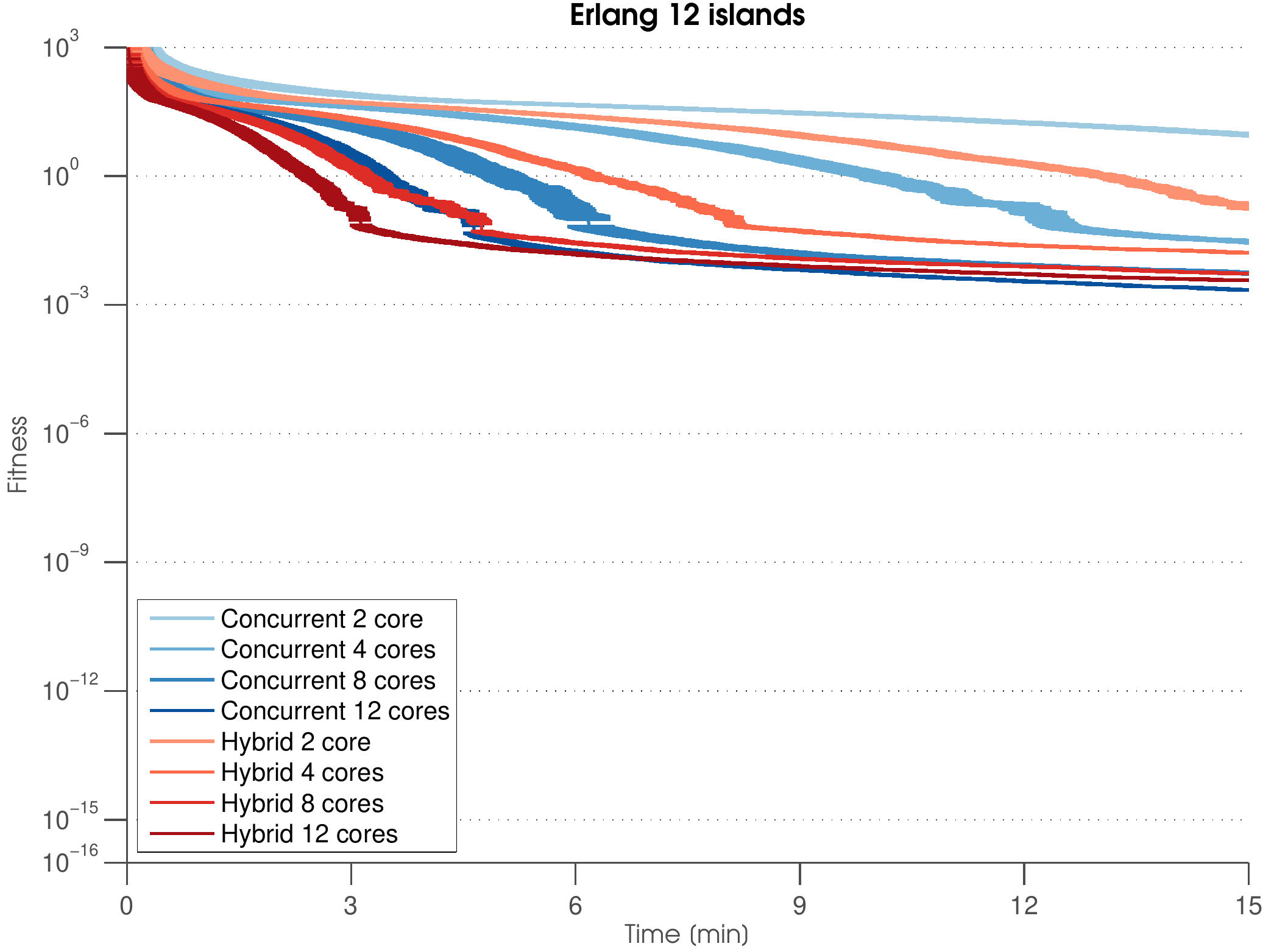}  
    
    \bigskip
    
    \includegraphics[width=\textwidth]{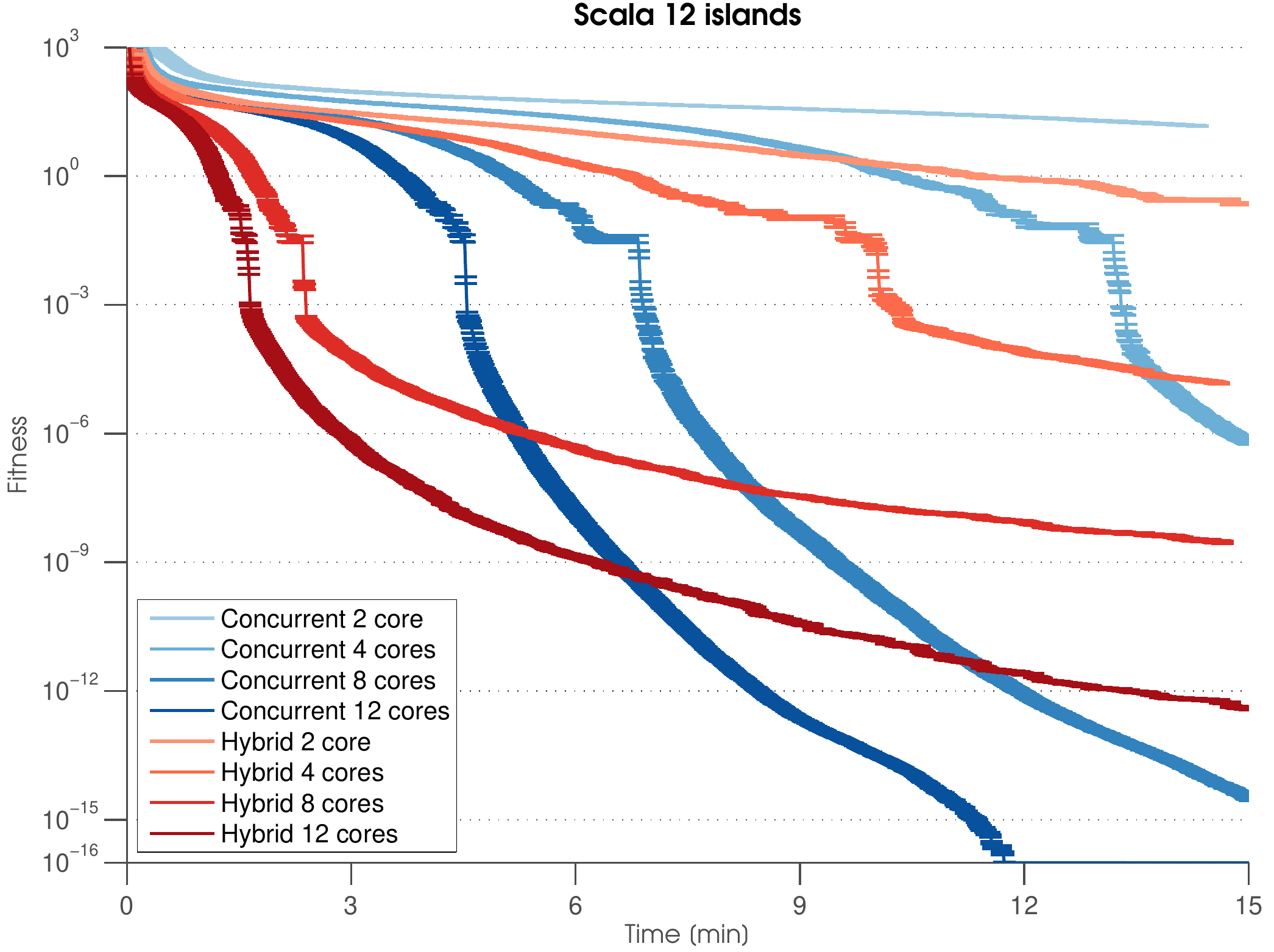}  
    \caption{Best fitness ever over time, depending on the model and the number of cores. Stripes indicate 95\% confidence intervals. ~$10^{-16}$ constant has been added to the fitness values to visualize the global optimum.}
    \label{fig:bestFitness}
\end{figure}

\paragraph{Results}

We examined our models under two criteria: how well the algorithm works and how fast the meeting mechanism is.

We assessed the quality of the algorithm by recording: the fitness of the best solution found so far at any given time on any island (see Fig. \ref{fig:bestFitness}).

\begin{figure}[!h]
    \centering 
    \includegraphics[width=\textwidth]{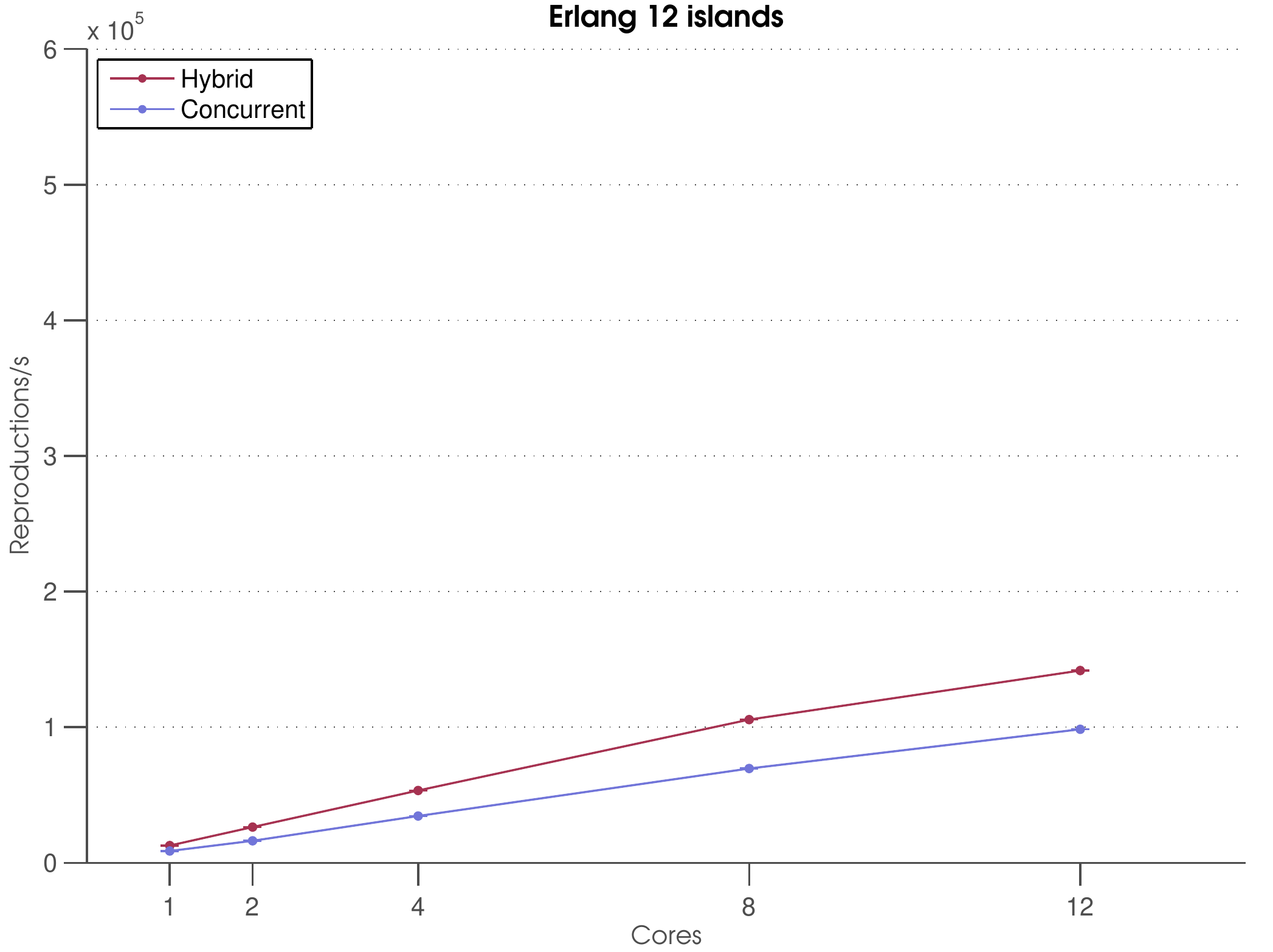}  
    \includegraphics[width=\textwidth]{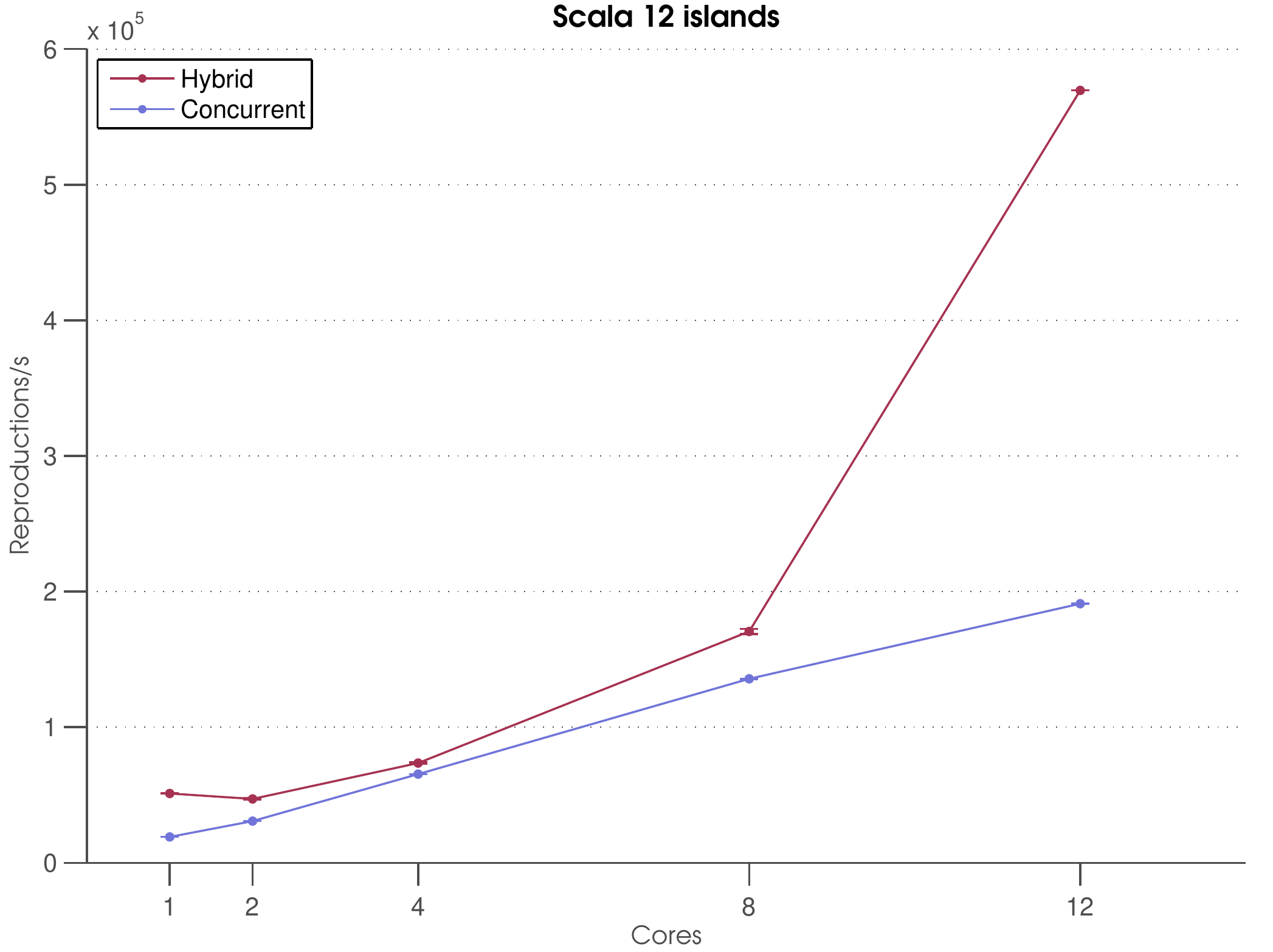}    
    \caption{The relation of the number of reproductions per second to the number of cores, for each model and implementation, along with 95\% confidence intervals.}
    \label{fig:reproduction}
\end{figure}

We estimated the speed of the models by counting the amount of agent meetings performed in a unit of time. These numbers appeared to be proportionally related across different arenas. Therefore, we only consider below the amount of reproductions per second (see Fig. \ref{fig:reproduction}).
The number of reproductions may depend not only on the implementation and number of cores but also on the algorithm itself. Therefore, it is a useful metric of speed when comparing the same model but with different implementations and numbers of cores. However, the number of reproductions relates to the number of fitness function evaluations, it is also a metric of efficiency between the alternate models.

\begin{figure}[!h]
    \centering 
    \includegraphics[width=\textwidth]{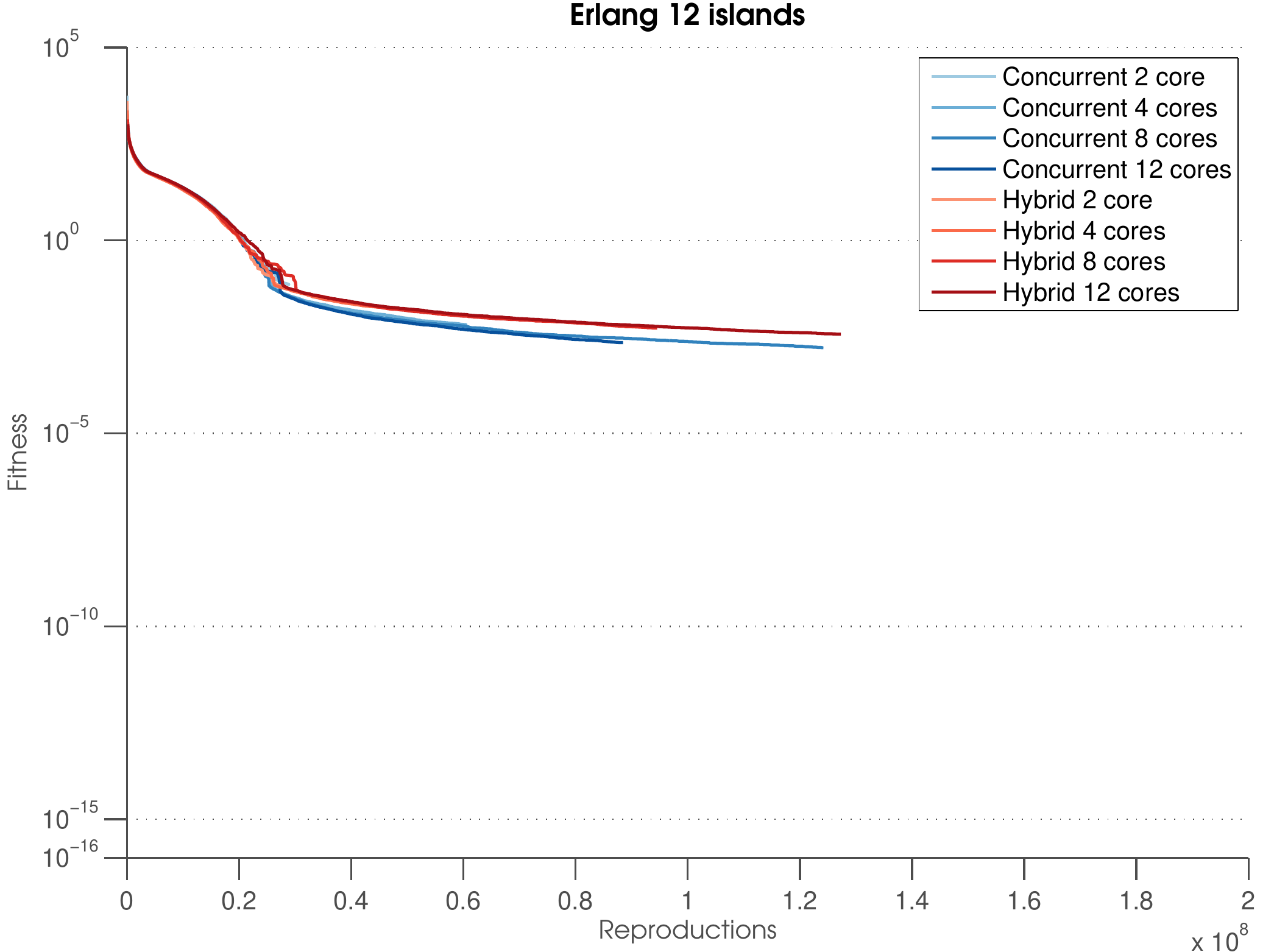}  
    
    \bigskip
    
    \includegraphics[width=\textwidth]{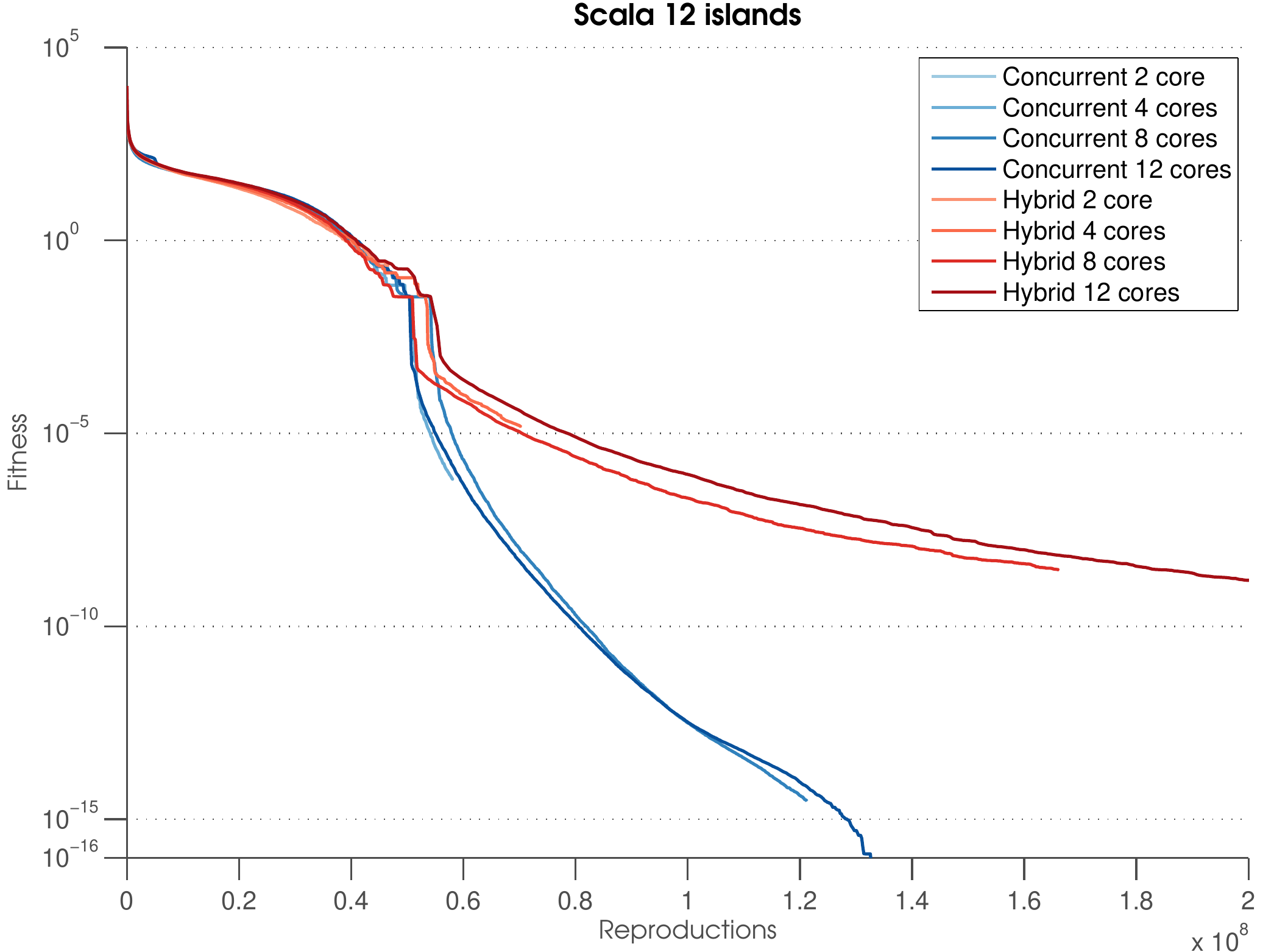}    
    \caption{Best fitness ever over \emph{the total number of reproductions}. Confidence intervals are the same as in Fig. \ref{fig:bestFitness} and have been omitted for clarity.}
    \label{fig:fit_repr}
\end{figure}

\paragraph{Discussion}

The results of the optimisation experiments are shown in Fig. \ref{fig:bestFitness}. Both models in both implementations improve when cores are added. The sequential versions in both Erlang and Scala behaved just like the hybrid models with 1 cores, minus some small communication overhead, so these results are omitted further on.

The results for the Erlang and Scala implementations are similar in the early stages of experiments. In the later stages, the Erlang version becomes much slower than the Scala one. However, a close-up on the Erlang results reveals that its characteristics are also similar to the Scala results, but over a larger timespan.

The Scala results show that the hybrid version is initially faster. However, the concurrent model takes over at some point and becomes much more effective than the hybrid model in the later stages. In fact, in the case of 12 cores, 100\% of the experiments with the concurrent model found the global optimum by the 12th minute of the experiments.

Another difference between the models is the number of reproduction happening every second (Fig. \ref{fig:reproduction}). In the case of the Erlang implementation, these numbers increase nearly linearly with the increase of nodes. 

In the Scala implementation, the concurrent version also scales linearly, but the hybrid version drastically improves when the number of cores equals the number of islands. Looking at it another way, the performance of the Scala hybrid model drastically declines when there is less cores than islands, which is not the case of the Erlang implementation. In that regard, Erlang real-time scheduling prove to be more efficient than naive JVM threading (thread per island).

In both implementations the number of reproduction per second is significantly smaller in the case of the concurrent version. The interpretation can be twofold: on one hand, they may indicate that the concurrent implementation is slow at processing agent meetings. On the other hand, the number of reproductions reflect the number of fitness evaluations. As the concurrent version still achieves better results, but with fewer fitness evaluations, it can be considered more efficient.

This observation becomes all the more evident when plotting the best fitness over the number of total reproductions instead of over time. Fig. \ref{fig:fit_repr} confirms that the concurrent and hybrid models are in fact two different algorithms. Increasing the number of cores simply increases the number of reproductions per second and therefore reduces the time to reach a given value. However, the concurrent version needs a much lower number of reproductions, and therefore less function evaluations, to reach a given value.

This difference in dynamics could be explained in the following way: in the hybrid version, agents in the population are effectively synchronized, in the sense that all fights and all reproductions in a step need to end for any agent to move on. In contrast, in the concurrent version fights happen independently of reproduction and the population evolves in a much more continuous way. Information spreads faster in the population and the solution can be found with fewer generations.

Therefore, as the concurrent version needs less function evaluation, we conjecture that it should perform even better compared to the hybrid one when faced with real-life problems, where the computation of the fitness function itself can take much time.

The difference in performance between the Erlang and Scala versions can have several causes.
First, the Erlang VM is usually less efficient than the Java VM when it comes to raw arithmetic. Both languages use 64 bit floating point numbers, though, so problems with numerical precision can be ruled out.
Second, both languages strongly differ in memory management. In Erlang, every process owns a separate stack. With a few exceptions, all messages need to be copied from the memory of one process to another, even if the data is immutable and could be sent by reference.
In contrast, Scala and Akka are based on the shared Java Memory Model \cite{manson2005java}, but offer different concurrency primitives in order to use message passing. Therefore, immutable data can be transferred within a VM as fast as in Java and as safely as in Erlang.
All in all, the Scala implementation incurs less overhead and the differences in characteristics between the algorithms are amplified. However, it is the insight from designing the algorithm for Erlang first that led us to that efficient implementation in Scala.

\section{Conclusions}

The massively-concurrent implementation of a evolutionary multi-agent system presented in this paper gives very promising results in terms of scalability and efficiency. Its asynchronous nature allows to better imitate the mechanisms observed in biological evolution, going beyond the classical approach of discrete generations and synchronous population changes.

We applied this algorithm to a popular optimization benchmark. The results of our experiments indicate that when many agents are involved, the concurrent model is significantly more efficient in terms of approaching the optimum versus number of fitness function evaluations. This result shows that this technique is very promising when the complexity of fitness function is high (e.g. in the case of solving inverse problems).

The key to achieve an efficient implementation was using Erlang and Scala technologies, in particular their features like lightweight processes and fast message passing concurrency. The Scala version appears to be more efficient, mainly because of a better memory management of the underlying library.

A further development of this method on modern multicore supercomputers \cite{kevin} seems a promising direction of research. Broader tests will also be performed in multicore systems consisting of a higher number of cores than examined here.

\section*{Acknowledgement}
The research presented in the paper was partially supported by the European Commission
FP7 through the project ParaPhrase: Parallel Patterns for Adaptive Heterogeneous Multicore Systems, under contract
no.: 288570 (http://paraphrase-ict.eu).
The research presented in this paper received 
partial financial support from AGH University of Science and Technology statutory project.
The research presented in the paper was conducted using PL-Grid Infrastructure (http://www.plgrid.pl/en).

\bibliographystyle{plain}
\bibliography{references}

\begin{thebibliography}{10}

\bibitem{Bellifemine2001}
Fabio Bellifemine, Agostino Poggi, and Giovanni Rimassa.
\newblock {JADE}: {A} {FIPA2000} {C}ompliant {A}gent {D}evelopment
  {E}nvironment.
\newblock In {\em Proceedings of the Fifth International Conference on
  Autonomous Agents}, AGENTS '01, pages 216--217, New York, NY, USA, 2001. ACM.

\bibitem{ker}
A.~Byrski, R.~Dre\.zewski, L.~Siwik, and M.~Kisiel-Dorohinicki.
\newblock Evolutionary multi-agent systems.
\newblock {\em The Knowledge Engineering Review}, 30:171--186, 2015.

\bibitem{korczynski}
A.~Byrski, W.~Korczy\'nski, and M.~Kisiel-Dorohinicki.
\newblock Memetic multi-agent computing in difficult continuous optimisation.
\newblock In {\em Advanced Methods and Technologies for Agent and Multi-Agent
  Systems}, pages 181--190. IOS Press, 2013.

\bibitem{biul1}
A.~Byrski, R.~Schaefer, and M.~Smo\l{}ka.
\newblock Asymptotic guarantee of success for multi-agent memetic systems.
\newblock {\em Bulletin of the Polish Academy of Sciences---Technical
  Sciences}, 61(1), 2013.

\bibitem{csci2013}
Aleksander Byrski.
\newblock Tuning of agent-based computing.
\newblock {\em Computer Science (AGH)}, 14(3):491, 2013.

\bibitem{facilitating}
Aleksander Byrski and Marek Kisiel-Dorohinicki.
\newblock Agent-based model and computing environment facilitating the
  development of distributed computational intelligence systems.
\newblock In Gabrielle Allen, Jarosław Nabrzyski, Edward Seidel, GeertDick van
  Albada, Jack Dongarra, and PeterM.A. Sloot, editors, {\em Computational
  Science – ICCS 2009}, volume 5545 of {\em Lecture Notes in Computer
  Science}, pages 865--874. Springer Berlin Heidelberg, 2009.

\bibitem{cecmoj}
Aleksander Byrski and Robert Schaefer.
\newblock Formal model for agent-based asynchronous evolutionary computation.
\newblock In {\em Proceedings of the {IEEE} Congress on Evolutionary
  Computation, {CEC} 2009, Trondheim, Norway, 18-21 May, 2009}, pages 78--85.
  {IEEE}, 2009.

\bibitem{cantu-paz1998survey}
E.~Cant{\'u}-Paz.
\newblock A survey of parallel genetic algorithms.
\newblock {\em Calculateurs Paralleles, Reseaux et Systems Repartis},
  10(2):141--171, 1998.

\bibitem{kc}
K.~Cetnarowicz, M.~Kisiel-Dorohinicki, and E.~Nawarecki.
\newblock The application of evolution process in multi-agent world ({MAW}) to
  the prediction system.
\newblock In M.~Tokoro, editor, {\em Proc. of the 2nd Int. Conf. on Multi-Agent
  Systems (ICMAS'96)}. AAAI Press, 1996.

\bibitem{DiazEarle2013}
Á.F. Díaz, C.B. Earle, and L-Å Fredlund.
\newblock e{J}ason: {A}n {I}mplementation of {J}ason in {E}rlang.
\newblock In Mehdi Dastani, J.F. Hübner, and Brian Logan, editors, {\em
  Programming Multi-Agent Systems}, volume 7837 of {\em Lecture Notes in
  Computer Science}, pages 1--16. Springer Berlin Heidelberg, 2013.

\bibitem{di2003exat}
Antonella Di~Stefano and Corrado Santoro.
\newblock {eXAT: an Experimental Tool for Programming Multi-Agent Systems in
  Erlang}.
\newblock In {\em WOA}, pages 121--127, 2003.

\bibitem{FrankSteffenRaubal2001}
Andrew~U. Frank, Steffen Bittner, and Martin Raubal.
\newblock Spatial and cognitive simulation with multi-agent systems.
\newblock In DanielR. Montello, editor, {\em Spatial Information Theory},
  volume 2205 of {\em Lecture Notes in Computer Science}, pages 124--139.
  Springer Berlin Heidelberg, 2001.

\bibitem{goldberg}
D.~E. Goldberg.
\newblock {\em Genetic algorithms in search, optimization, and machine
  learning.}
\newblock Addison-Wesley, 1989.

\bibitem{GrigoreCollier2011}
C.~Grigore and R.~Collier.
\newblock Supporting agent systems in the programming language.
\newblock In {\em Web Intelligence and Intelligent Agent Technology (WI-IAT),
  2011 IEEE/WIC/ACM International Conference on}, volume~3, pages 9--12, Aug
  2011.

\bibitem{Gutknecht2001}
Olivier Gutknecht and Jacques Ferber.
\newblock {The madkit agent platform architecture}.
\newblock {\em Infrastructure for Agents, Multi-Agent Systems, and Scalable
  Multi-Agent Systems}, 2001.

\bibitem{kevin}
K.~Hammond, M.~Aldinucci, C.~Brown, F.~Cesarini, M.~Danelutto,
  H.~Gonzalez-Velez, P.~Kilpatrick, R.~Keller, M.~Rossbory, and G.~Shainer.
\newblock The paraphrase project: Parallel patterns for adaptive heterogeneous
  multicore systems.
\newblock In {\em FMCO: 10th International Symposium on Formal Methods for
  Components and Objects-Revised Selected Papers}, volume 7542, pages 218--236.
  Springer LNCS, 2013.

\bibitem{JenSycWoo:AAMAS98}
N.~R. Jennings, K.~Sycara, and M.~Wooldridge.
\newblock A roadmap of agent research and development.
\newblock {\em Journal of Autonomous Agents and Multi-Agent Systems},
  1(1):7--38, 1998.

\bibitem{DBLP:conf/ecms/KolybaczKLBK13}
Magdalena Kolybacz, Michal Kowol, Lukasz Lesniak, Aleksander Byrski, and Marek
  Kisiel{-}Dorohinicki.
\newblock Efficiency of memetic and evolutionary computing in combinatorial
  optimisation.
\newblock In Webj{\o}rn Rekdalsbakken, Robin~T. Bye, and Houxiang Zhang,
  editors, {\em Proceedings of the 27th European Conference on Modelling and
  Simulation, {ECMS} 2013, {\AA}lesund, Norway, May 27-30, 2013}, pages
  525--531. European Council for Modeling and Simulation, 2013.

\bibitem{DBLP:conf/iccS/KowolBK14}
Michal Kowol, Aleksander Byrski, and Marek Kisiel{-}Dorohinicki.
\newblock Agent-based evolutionary computing for difficult discrete problems.
\newblock In David Abramson, Michael Lees, Valeria~V. Krzhizhanovskaya, Jack
  Dongarra, and Peter M.~A. Sloot, editors, {\em Proceedings of the
  International Conference on Computational Science, {ICCS} 2014, Cairns,
  Queensland, Australia, 10-12 June, 2014}, volume~29 of {\em Procedia Computer
  Science}, pages 1039--1047. Elsevier, 2014.

\bibitem{Krzywicki2014}
D.~Krzywicki, Ł. Faber, A.~Byrski, and M.~Kisiel-Dorohinicki.
\newblock Computing agents for decision support systems.
\newblock {\em Future Generation Computer Systems}, 37:390--400, 2014.

\bibitem{Krzywicki20141068}
D.~Krzywicki, J.~Stypka, P.~Anielski, Ł. Faber, W.~Turek, A.~Byrski, and
  M.~Kisiel-Dorohinicki.
\newblock Generation-free agent-based evolutionary computing.
\newblock {\em Procedia Computer Science}, 29(0):1068 -- 1077, 2014.
\newblock 2014 International Conference on Computational Science.

\bibitem{ManateMunteanu2013}
B.~Manate, V.I. Munteanu, and T.-F. Fortis.
\newblock Towards a scalable multi-agent architecture for managing iot data.
\newblock In {\em P2P, Parallel, Grid, Cloud and Internet Computing (3PGCIC),
  2013 Eighth International Conference on}, pages 270--275, Oct 2013.

\bibitem{manson2005java}
Jeremy Manson, William Pugh, and Sarita~V. Adve.
\newblock The java memory model.
\newblock {\em SIGPLAN Not.}, 40(1):378--391, January 2005.

\bibitem{michalewicz}
Z.~Michalewicz.
\newblock {\em Genetic Algorithms Plus Data Structures Equals Evolution
  Programs}.
\newblock Springer-Verlag New York, Inc., Secaucus, NJ, USA, 1994.

\bibitem{Piotrowski2013}
M.~Piotrowski and W.~Turek.
\newblock {Software Agents Mobility Using Process Migration Mechanism in
  Distributed Erlang}.
\newblock In {\em Proceedings of the Twelfth ACM SIGPLAN Workshop on Erlang},
  Erlang '13, pages 43--50, New York, NY, USA, 2013. ACM.

\bibitem{pisarskirugala}
Sebastian Pisarski, Adam Rugała, Aleksander Byrski, and Marek
  Kisiel-Dorohinicki.
\newblock Evolutionary multi-agent system in hard benchmark continuous
  optimisation.
\newblock In AnnaI. Esparcia-Alcázar, editor, {\em Applications of
  Evolutionary Computation}, volume 7835 of {\em Lecture Notes in Computer
  Science}, pages 132--141. Springer Berlin Heidelberg, 2013.

\bibitem{Pokahr2013}
Alexander Pokahr, Lars Braubach, and Kai Jander.
\newblock The jadex project: Programming model.
\newblock In Maria Ganzha and Lakhmi~C. Jain, editors, {\em Multiagent Systems
  and Applications}, volume~45 of {\em Intelligent Systems Reference Library},
  pages 21--53. Springer Berlin Heidelberg, 2013.

\bibitem{polnik}
M.~Polnik, M.~Kumiega, and A.~Byrski.
\newblock Agent-based optimization of advisory strategy parameters.
\newblock {\em Journal of Telecommunications and Information Technology},
  2:54--55, 2013.

\bibitem{hgscamwa}
Robert Schaefer, Aleksander Byrski, Joanna Kolodziej, and Maciej Smolka.
\newblock An agent-based model of hierarchic genetic search.
\newblock {\em Comput. Math. Appl.}, 64(12):3763--3776, December 2012.

\bibitem{schwefel}
Hans-Paul Schwefel and Gunter Rudolph.
\newblock Contemporary evolution strategies.
\newblock In {\em European Conference on Artificial Life}, pages 893--907,
  1995.

\bibitem{turek2013erlang}
Wojciech Turek.
\newblock Erlang as a high performance software agent platform.
\newblock {\em Advanced Methods and Technologies for Agent and Multi-Agent
  Systems}, 252:21, 2013.

\bibitem{vose}
M.~Vose.
\newblock {\em The Simple Genetic Algorithm: Foundations and Theory}.
\newblock MIT Press, Cambridge, MA, USA, 1998.

\bibitem{wolpert}
D.H. Wolpert and W.G. Macready.
\newblock No free lunch theorems for optimization.
\newblock {\em IEEE Transactions on Evolutionary Computation}, 67(1), 1997.

\bibitem{DBLP:journals/aghcs/WrobelTPB13}
Krzysztof Wrobel, Pawel Torba, Maciej Paszynski, and Aleksander Byrski.
\newblock Evolutionary multi-agent computing in inverse problems.
\newblock {\em Computer Science {(AGH)}}, 14(3):367--384, 2013.

\end{thebibliography}


\end{document}